\documentclass[twocolumn,aps,amsmath,nofootinbib,superscriptaddress]{revtex4}

\usepackage{graphicx}
\usepackage{bm}
\usepackage{epsfig}

  \newcount\hour \newcount\minute
  \hour=\time \divide \hour by 60
  \minute=\time
  \newcommand{\mydate}{\ \today \ - \number\hour :\ifnum \minute<10 0\fi 
\number\minute}

\def\bnslash{\bar n\!\!\!\slash}

\def\OMIT#1{}

\newcommand{\nn}{\nonumber} 

\newcommand{\bn}{{\bar n}}
\newcommand{\bea}{\begin{eqnarray}}
\newcommand{\eea}{\end{eqnarray}}

\newcommand{\mcdot}{\!\cdot\!}

\newcommand{\SCETa}{\mbox{${\rm SCET}_{\rm I}$ }}
\newcommand{\SCETb}{\mbox{${\rm SCET}_{\rm II}$ }}

\def\lqcd{\Lambda_{\rm QCD}}

\begin{document}


\preprint{ \hbox{MIT-CTP 3469} \hbox{CMU-HEP-04-01} \hbox{CALT-68-2475}
 \hbox{hep-ph/040xxxx}  }

\title{\boldmath 
On Differences Between SCET and QCDF for $B\to\pi\pi$ Decays \\
}
\author{Christian W.~Bauer}
\affiliation{California Institute of Technology, Pasadena, CA 91125}
\author{Dan Pirjol}
\affiliation{Center for Theoretical Physics, Massachusetts Institute of
  Technology, Cambridge, MA 02139}
\author{Ira Z.~Rothstein}
\affiliation{Department of Physics, Carnegie Mellon University,
    Pittsburgh, PA 15213  \vspace{0.3cm}}
\author{Iain W. Stewart\vspace{0.4cm}}
\affiliation{Center for Theoretical Physics, Massachusetts Institute of
  Technology, Cambridge, MA 02139}
\begin{abstract}
  
  We give a detailed description of the differences between the factorization
  and results derived from SCET and QCDF for decays $B \to M_1 M_2$. This serves
  as a reply to the comment about our work
  {\it $B \to M_1 M_2$: Factorization,
    charming penguins, strong phases, and polarization}~\cite{bprs} 
  made by the authors in~\cite{BBNSComment}. We disagree with their criticisms.

\end{abstract}

\maketitle

In~\cite{bprs} we derived a factorization formula for exclusive $B$ decays to
two light mesons using the soft collinear effective theory (SCET)~\cite{SCET}.
Recently, Beneke, Buchalla, Neubert and Sachrajda posted a comment about our
work~\cite{BBNSComment}, and compared it with their QCDF (QCD factorization)
approach~\cite{BBNS}. In this paper we compare results and reply to
their comments~\cite{BBNSComment}.

For easy reference, we summarize a few points made in Ref.~\cite{bprs} that
disagree with Ref.~\cite{BBNS}. We found that: i) a proper separation of scales
$Q^2\gg E_\pi\Lambda\gg \Lambda^2$ in the factorization theorem are different in
SCET and QCDF, and can be regarded as a formal disagreement if desired ($Q=
m_b,E_\pi$); ii) only a subset of $\alpha_s(m_b)$ corrections are currently
known, so results for these corrections are formally incomplete; iii) certain
amplitudes are sensitive to the treatment of $m_c$, with parametrically large
$\sim v$ contributions from $c\bar c$ in the NRQCD region where $v$ is the
velocity power counting parameter; iv) current $B\to \pi\pi$ data analyzed at LO
in SCET supports values for the parameters $\zeta^{B\pi}\sim \zeta_J^{B\pi}$, in
disagreement with numerical inputs adopted in QCDF.  We also emphasized that
$\Lambda/m_b$ power corrections need to be of natural size to support model
independent phenomenology and verify that the expansion converges, concepts
which are sometimes relaxed in QCDF phenomenological analyses.  If power
corrections change the LO values of $\zeta^{B\pi}$ and $\zeta_J^{B\pi}$
substantially then this would indicate that the power expansion is not
converging and a different expansion would be needed if model independent
results are desired.

We take this opportunity to also comment on results from Ref.~\cite{bprs} where
we found agreement with points made in Ref.~\cite{BBNS}. The original starting
idea is the same, that factorization theorems for these decays should be derived
by making a systematic expansion of QCD in $\Lambda_{\rm QCD}/m_b$, where the
terms in this expansion are model independent and unique. Earlier discussion of
QCD based factorization methods for nonleptonic decays can be found
in~\cite{Brodsky,earlier}. We agree on the scaling in $\Lambda/m_b$ for the LO
amplitudes.  There is agreement that input on non-perturbative functions can be
obtained from $B\to \pi$ form factors and LO light-cone meson distribution
functions $\phi_\pi(x)$ and $\phi_B(k^+)$. We also agree that at LO
factorization occurs for amplitudes from light quark penguin loops, as well as
tree, and color-suppressed diagrams. Finally, the set of the one-loop hard
corrections computed in Ref.~\cite{BBNS} determine the Wilson coefficients of
the $Q^{(0)}_i$ operators~\cite{chay}.

It is also worth emphasizing that the scope of our two works was different. In
Refs.~\cite{BBNS,BN} factorization theorems were proposed based on the study of
the IR singularities of lowest order diagrams in perturbation theory.  Input
parameters are taken from QCD sum rules. Certain power suppressed contributions
were also included, although the factorization was not extended to this order;
as a result, some of these corrections are IR divergent, and cutoffs were used
for numerical estimates. In this way the authors of Refs.~\cite{BBNS,BN} were
able to make predictions for many modes, which however depend on model dependent
input.  In contrast, in~\cite{SCET} we used operators in SCET to separate the
long and short distance physics to all orders in $\alpha_s$.  The decay
amplitudes factor, with the long distance physics given by a few universal
hadronic parameters.  Predictive power was shown to be retained even when
$\alpha_s(\sqrt{E \Lambda})$ effects are summed to all orders.  We then used
data to determine these LO hadronic parameters and obtained a prediction for the
heavy to light form factor $f_+(0)$. The factorization theorem also gives a
model independent determination of the weak phase $\gamma$ (or $\alpha$) using
$Br(B\to\pi^0\pi^0)$ as input, but not $C_{\pi^0\pi^0}$~\cite{gamma}.

We organize the remainder of this paper by the four points raised
in~\cite{BBNSComment} which we disagree with: 1) that the SCET
result~\cite{bprs} is formally equivalent to the QCDF result~\cite{BBNS,BN} in
all respects, 2) that there is little benefit to avoiding the perturbative
expansion at the scale $\mu = \sqrt{\lqcd\, m_b}$, 3) that operators containing
a charm quark pair are perturbatively calculable, with corrections suppressed by
$\Lambda/m_b$ regardless of how the scale $m_c$ is treated, and 4) that our
phenomenological analysis of recent $B\to\pi\pi$ data, which disfavors certain
QCDF input parameters, is flawed because it omits ``known'' perturbative and
power suppressed contributions. A section is devoted to each of these topics.

\section{Formal Comparison}

In SCET the separation of scales $Q^2\gg E_\pi\Lambda\gg \Lambda^2$ can be
achieved by matching QCD onto a theory called \SCETa to integrate out
$Q^2$~\cite{bfprs}, and then matching \SCETa onto a final theory \SCETb to
integrate out the scale $E_\pi\Lambda$~\cite{bps4}.  Performing the first step
for $A_{\pi^+\pi^-}=A(\bar B\to\pi^+\pi^-)$ at LO in the power counting we 
found~\cite{bprs}
\begin{eqnarray} \label{Apipi}
 A_{\pi^+\pi^-} \!\!&=&\!\! 
N \bigg\{
   f_{\pi}\! \int\!\!du\, dz\,
    T_{1\!J}(u,z) \zeta^{B\pi}_{J}(z) \phi^{\pi}(u) 
    \\
 &&\hspace{-0.9cm}
   +\,  \zeta^{B\pi}\, f_{\pi}\!\! \int\!\! du\, T_{1\zeta}(u) \phi^{\pi}(u)
  \bigg\}  + \lambda_c^{(f)} A_{c\bar
    c}^{\pi\pi} \,, \nn
\end{eqnarray}
where $N=G_F m_B^2/\sqrt{2}$, the $T_i$'s capture hard $\alpha_s(m_b)$
contributions, and $\zeta_J^{B\pi}$, $\zeta^{B\pi}$ depend on the
$\sqrt{E_\pi\Lambda}$ and $\Lambda$ scales.  The analogous expression for the
$B\to\pi$ form factor is
\begin{eqnarray}  \label{fplus}
 f_+(0)\!\!&=&\!\! T^{(+)}\,\zeta^{B\pi} + 
  \int\!\! dz \, \hat C_J^{(+)}(z)\, \zeta_J^{B\pi}(z)  \,,
\end{eqnarray}
from which we observed that both observables depend on the same universal
$\zeta_J^{B\pi}(z)$ and $\zeta^{B\pi}$. Currently $T_{1\zeta}(u)$ is known at
${\cal O}(\alpha_s(m_b))$~\cite{BBNS}, {\em but} $T_{1J}(u)$ {\em is not}, hence
our statement that the calculation of the hard $\alpha_s(m_b)$ corrections are
incomplete.  In the notation in~\cite{bprs} the one-loop matching for the Wilson
coefficients $b_i^{(f)}(u,z)$ are missing.  We do not believe that these facts
are disputed in~\cite{BBNSComment}.  The amplitude $A_{c\bar c}^{\pi\pi}$
denotes long-distance $c\bar c$ contributions which we take up in a separate
section.  Short distance $c\bar c$ can contribute to $T_{1J}$ and $T_{1\zeta}$.

\begin{figure}[t!]
  \centerline{ \mbox{\epsfysize=2truecm \hbox{\epsfbox{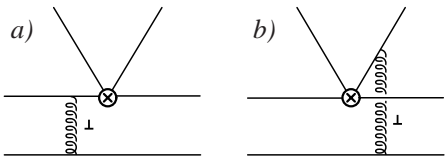}} } }
  \vskip-0.1cm {\caption[1]{Examples of QCD diagrams. In Eq.~(\ref{Apipi}) a)
      and b) both contribute to the $\zeta_J^{B\pi}(z)$ term in the
      factorization theorem, since the $\perp$-gluon in b) decouples from the
      upward going pion and can be Fierzed into the other quark bilinear. In
      Eq.~(\ref{ApipiQCDF}) graph a) contributes to the form factor term, and b)
      contributes to the hard-scattering term.}
\label{fig_compare} }
\vskip -0.2cm
\end{figure}

In Ref.~\cite{BBNS} the QCDF factorization formula was
\begin{eqnarray} \label{ApipiQCDF}
 A_{\pi^+\pi^-} \!\!&=&\!\! 
  N \: f_+(0) f_\pi \!\! \int\!\! du\, T_{\rm I}(u) \phi^{\pi}(u) \\
  &&\hspace{-1.4cm}
   +  N\, f_\pi^2 f_B
    \int\!\!du\, dx\, dk^+
    T_{\rm II}(u,x,k^+)  \phi_{\pi}(u) \phi_{\pi}(x) \phi_B(k^+)
  \,. \nn
\end{eqnarray}
Both of the scales $m_b$ and $\sqrt{E\Lambda}$ are treated perturbatively in
$T_{\rm II}$, so this result does not formally distinguish between these scales.
This makes it impossible to work to all orders in $\alpha_s$ at the
$\mu\sim\sqrt{E\Lambda}$ scale. It is also not possible to sum logarithms
between $m_b$ and $\sqrt{E_\pi\Lambda}$ without further factorization of $T_{\rm
  II}$.  Fig.~\ref{fig_compare} gives an example of how individual diagrams are
treated differently in (\ref{Apipi}) and (\ref{ApipiQCDF}) as explained in the
caption.
The result in Eq.(\ref{Apipi}) separates out the hard contributions $\sim Q^2$
regardless of the sensitivity to smaller scales $\sqrt{E_\pi\Lambda}$ and
$\Lambda$.

Separating the $E_\pi\Lambda \gg \Lambda^2$ scales is more complicated.  For
$\zeta_J^{B\pi}(z)$ we found 
\begin{eqnarray} \label{zetaJ}
\zeta_J^{B\pi}(z) = f_\pi f_B \int\!\! dk_+\! \int\!\! dx\:
  J(z,x,k_+) \phi_\pi(x) \phi_B^+(k_+) \,,
\end{eqnarray}
where the jet function $J$ starts at ${\cal O}(\alpha_s(\sqrt{E_\pi\Lambda}))$
and captures all corrections at this scale. (Multiplicative $\alpha_s(m_b)$
corrections in $f_B$ can be moved into $T_{1J}$ if desired.) The jet function
$J$ is now known at one-loop order~\cite{BH}.  As first discussed
in~\cite{chay}, using SCET one can write $T_{\rm II}(u,x,k^+)= \int\!\! dz
[T_{1J}(u,z)-\hat C_J^{(+)}(z)] J(z,x,k^+)$; a point mentioned
in~\cite{BBNSComment} on which we agree. A full disentangling of the scales
$E_\pi\Lambda$ and $\Lambda^2$ in $\zeta^{B\pi}$ is still being
debated~\cite{bps4,bffinite,messenger,bds,LN}.  In \SCETa the diagrams that
define $\zeta^{B\pi}$ involve the exchange of at least one hard-collinear
gluon~\cite{bps4}, just as they do for $\zeta_J^{B\pi}$, leading to the
expectation that this parameter should also start at ${\cal
  O}(\alpha_s(\sqrt{E_\pi\Lambda}))$, and $\zeta_J^{B\pi}\sim \zeta^{B\pi}$. The
parameter counting in QCDF assumes an $\alpha_s(\sqrt{E_\pi\Lambda}))$ only for
the hard spectator contributions, and in our notation they have
$\zeta_J^{B\pi}\ll \zeta^{B\pi}$.  The theoretical issue that blurs the answer
to this question is that naively performing similar steps for $\zeta^{B\pi}$ to
those giving Eq.~(\ref{zetaJ}) give divergent convolution integrals, indicating
that this parameter is ``non-factorizable''~\cite{Brodsky,bf,bps4,bffinite}.  A
complete understanding of the separation of the $\sqrt{E_\pi\Lambda}$ and
$\Lambda$ scales is necessary to sum all logs below the hard collinear scale.
For the contribution in Eq.~(\ref{zetaJ}) these logarithms have been resumed in
Ref.~\cite{HBLN}, and the running for the other part was studied in~\cite{LN}.
It has been argued~\cite{LN} that the $\alpha_s(\sqrt{E_\pi\Lambda})$ is absent
in $\zeta^{B\pi}$, however this relies on the conjecture that diagrams
containing a soft-collinear messenger mode~\cite{messenger} in the theory below
the scale $\sqrt{E_\pi\Lambda}$ cancels all endpoint singularities.
 
\section{Perturbations theory in $\alpha_s(\sqrt{E\Lambda})$}

In~\cite{bprs} we showed that an expansion in $\alpha_s(\sqrt{E_\pi\Lambda })$
is not required to obtain predictions from factorization in nonleptonic decays.
In other words Eq.~(\ref{Apipi}) has nontrivial implications even without using
a perturbative expansion for $J$ in Eq.~(\ref{zetaJ}).  Since
$\sqrt{E_\pi\Lambda} \sim 1.0-1.6$ GeV, we believe that it is useful to consider
what predictions can be made by avoiding this expansion.

In our opinion the advantage of using Eq.~(\ref{zetaJ}) with an expansion of $J$
in $\alpha_s(\sqrt{E_\pi\Lambda})$ is that it reduces $\zeta_J^{BM}(z)$, which
depends on the choice of $M$, to two more universal nonperturbative functions
$\phi_B^+(k^+)$ and $\phi_M(x)$. Thus, the number of unknowns decreases in
global analyses involving different light mesons $M$, and processes like
$\gamma^*\gamma\to M$ and $B\to \gamma e\bar \nu$~\cite{KPY} can be used to
provide additional nonperturbative information.  This is true both at tree level
in $\alpha_s(m_b)$ and at one-loop in $\alpha_s(m_b)$.

There were two main critiques raised in~\cite{BBNSComment}. First, it was argued
that there is no point in working to all orders in $\alpha_s(\sqrt{E_\pi\Lambda
})$ since perturbation theory in this quantity is reasonably well behaved, and
second that if one works to all orders in $\alpha_s(\sqrt{E_\pi\Lambda })$ then
the factorization theorem looses all predictive power beyond tree-level in
$\alpha_s(m_b)$.

Regarding the first point, it is always better to use less theoretical
assumptions if predictions at the level of precision currently achievable
actually do not rely on making this expansion.  The perturbative expansion at
the scale $\sqrt{m_b \Lambda_{\rm QCD}}$ may in fact converge (recent evidence
has been given by the one-loop calculation in~\cite{BH} and also relations
between $B\to D\eta$ and $B\to D\eta'$ decays~\cite{BDeta}), however if a
prediction is independent of this expansion then there is no need to rely on it.
Regarding the second point, it is true that beyond LO in $\alpha_s(m_b)$ the
functional dependence of $\zeta_J^{B\pi}(z)$ is required, instead of just the
number $\zeta_J^{B\pi} \equiv \int\!\!  dz\: \zeta_J^{B\pi}(z)$. However, the
same function determines the $B\to \pi$ form factors.  Note that when $\alpha_s$
corrections are included in Eq.~(\ref{ApipiQCDF}) the functional form of the $B$
meson wave function $\phi_B(k^+)$ is required. So in either case one has moments
of one unknown function.  We believe that the most important advantage of the
$\alpha_s(\sqrt{m_b \Lambda})$ expansion is the universality of the
nonperturbative functions mentioned above, rather than the change in how
$\alpha_s(m_b)$ corrections are included.

\section{Charm loops and $A_{c\bar c}$}

The size of charm loop contributions to $B\to\pi\pi$ are important. Certain
charm loop contributions are from hard ($\sim\! m_b$) momenta and there is broad
agreement~\cite{BSS,BBNS,chay,bprs} that these effects can be computed in
perturbation theory. In Eq.~(\ref{Apipi}) they enter in both $T_{1\zeta}$ and
$T_{1J}$. The point being debated is the parametric scaling of non-perturbative
contributions from penguin charm quark loops (so-called charming
penguins~\cite{cpens}), denoted by $A_{c\bar c}^{\pi\pi}$ in Eq.~(\ref{Apipi}).

For the parametric scaling two useful limits are
\begin{eqnarray} \label{limits}
  \mbox{i):  } && \frac{m_c}{m_b} \ll 1 \,, 
     \qquad \frac{\Lambda}{m_c}\ll 1 \\
 \mbox{ii):  } && \frac{m_c}{m_b} \sim {\cal O}(1) \,, 
     \qquad \frac{\Lambda}{m_c}\ll 1 
   \nn\,. 
\end{eqnarray}
In~\cite{BBNSComment} this corresponds to the limits $m_b \rightarrow \infty$
with $m_c$ fixed and $m_b,m_c\rightarrow \infty$ with $m_c/m_b$ fixed
respectively, however we believe the description in Eq.~(\ref{limits}) makes
aspects of the expansion more clear. For example, the charm quark power counting
will not be identical to that for light quarks unless $m_c\sim \Lambda$, which
is not realized in nature.  

In Ref.~\cite{bprs} we focused on nonperturbative contributions from $c\bar c$
in the NRQCD region and found that these contributions are only suppressed by
$v$. Note that these $c\bar c$'s can still have a total energy $\sim m_b$ as
long as their relative velocity $\sim v$ is small.\footnote{Other charm modes
  besides the ones considered here could also contribute to $A_{c\bar
    c}^{\pi\pi}$. The consideration of the charm modes above is sufficient to
  demonstrate the scaling of this nonperturbative contribution.}  Here $v$ is a
place holder for a nonperturbative matrix element which is parametrically of
this size. In charmonium $m_c v\sim 800\,{\rm MeV}$ and $m_c v^2\sim 400\,{\rm
  MeV}$~\cite{FLR}, so if we identify one of these scales with $\Lambda_{\rm
  QCD}$ the $v$ suppression becomes either a $\Lambda_{\rm QCD}/m_c$ or a
$\sqrt{\Lambda_{\rm QCD}/m_c}$ suppression. We do not think that the case $m_c
v^2\gg \Lambda_{\rm QCD}$ is physically relevant for charm quarks in QCD.  In
either case {\em if we expand} in $\Lambda/m_c$ these nonperturbative
contributions do not enter the LO $B\to \pi\pi$ factorization theorem.  However,
in practice $v\sim 0.5$ so these corrections are numerically large compared to
$\Lambda/m_b\sim 0.1$ power corrections, and can spoil the power expansion.

In Refs.~\cite{BBNS,FH,BBNSComment} it was argued that the NRQCD $c\bar c$
region does not require special treatment due to quark-hadron duality, with
smearing from the $q^2 =\bar x m_b^2$ of the gluon which the charm annihilate
into ($0\le q^2\le m_b^2$). Using duality in this sense requires an inclusive
hadronic final state, to make it possible for there to be a cancellation of
infrared divergences between the virtual and real diagrams to all orders in
$\alpha_s$.  For exclusive decays like $B\to\pi\pi$ one must instead prove a
factorization theorem to separate hard and infrared contributions.  In these
proofs one must consider the contributions from all possible momentum regions.

The arguments in~\cite{BBNSComment} assume that the size of the nonperturbative
$c\bar c$ terms can be estimated based on regions of phase space in $q^2$,
taking limits based on a factorization formula analogous to (\ref{ApipiQCDF}).
To the best of our knowledge it has never been proven that the NRQCD $c\bar c$
contributions factor in this way.  Intuitively we expect that they will not.
The gluons whose wavelength is $\sim\Lambda$ do not decouple from the charm pair
which are created and annihilated in the octet state.  Since the $c\bar c$
production and annihilation occur over a distance scale $\sim \Lambda_{\rm
  QCD}^{-1}$, the soft gluons radiated from energetic quarks produced from the
annihilation may not cancel.  This can lead to two types of Wilson lines,
$Y_n[n\mcdot A_{\rm soft}]$ and $Y_\bn[\bn\mcdot A_{\rm soft}]$ in the soft
$B$-matrix element. In this case the amplitude will involve a new
nonperturbative function which has a strong phase from the mechanism found
in~\cite{mps}, since the soft function carries information about the final state
through $n$ and $\bn$.

Even in the absence of a proof of factorization for $A_{c\bar c}^{\pi\pi}$, it
should however still be possible to determine its parametric dependence on
$m_c/m_b$, $v$, and $\Lambda_{\rm QCD}/m_b$ using operators in effective field
theories. We find
\begin{eqnarray} \label{Acc}
 \frac{A_{c\bar c}^{\pi\pi}}{A^{\pi\pi}_{LO}} \sim
  \alpha_s(2m_c) \: f\Big(\frac{2m_c}{m_b}\Big)\: v 
 \,,
\end{eqnarray}
in agreement with~\cite{bprs}. Eq.(\ref{Acc}) disagrees with the result
in~\cite{BBNSComment} since there is no $\Lambda$ in the numerator besides that
hidden in $v$.  Physically $f(2m_c/m_b)$ encodes the restriction of the charm
quarks to be produced with small relative velocity (rather than for example
back-to-back with energies $\sim m_b/2$). The factor of $v$ gives the remaining
suppression for the charm quarks to be non-perturbative in the NRQCD region.
Together these include all ``phase-space suppression'' factors,
which~\cite{BBNSComment} claims were missed in~\cite{bprs}. A derivation of
Eq.~(\ref{Acc}) is given in the Appendix.

\section{Numerical Values of $\zeta_J^{B\pi}$, $\zeta^{B\pi}$}
\label{sec:pheno}

\begin{figure}[t!]
  \centerline{ \mbox{\epsfysize=4.4truecm \hbox{\epsfbox{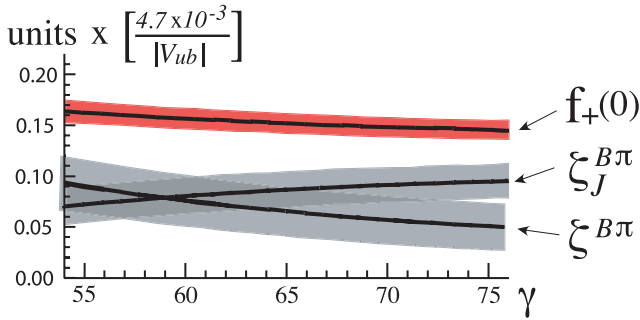}} } }
  \vskip-0.2cm {\caption[1]{An update of Fig.~5 in~\cite{bprs} giving model
      independent results for $\zeta^{B\pi}$, $\zeta_J^{B\pi}$, and the $B\to
      \pi$ form factor $f_+(q^2=0)$ as a function of $\gamma$ and $|V_{ub}|$.
      The shaded bands show the $1$-$\sigma$ errors propagated from the
      $B\to\pi\pi$ data.}
\label{fig:ff} }
\vskip -0.4cm
\end{figure}
In \cite{bprs} the parameters $\zeta$ and $\zeta_J$ were extracted from a LO
SCET analysis of the $B\to\pi\pi$ data. New data~\cite{pipidata} was presented
at ICHEP 2004, and we give here an update of the analysis of these parameters
from \cite{bprs}.  We compare our results with the most recent QCDF analysis
in~\cite{BN} and address the criticism in~\cite{BBNSComment}.

\begin{table*}[t!] 
\begin{ruledtabular}
\begin{tabular}{c|l|l|l|l}
 & QCDF default scenario & QCDF scenario 2 & ${\rm SCET}_{\rm LO}$, 
 $\langle x^{-1}\rangle_\pi=\{3.3,4.2\}$ & data \\
\hline
$T$ 
 & $0.354 - [0.037 + 0.019 i + \ldots]$ 
 & $0.358 - [0.033 + 0.017 i+ \ldots]$  
 & $1.19 \zeta^{B\pi}+ \{2.17, 2.44\} \zeta_J^{B\pi}$
 & $|T|=0.30 \pm 0.02$ \\
 & \hspace{0.2cm}  $+ (0.006 X_H)$ 
 &\hspace{0.2cm}  $+ (0.016 X_H)$  &   \\
$T_c$ 
 & $0.272 + [0.008+0.0i+ \ldots] $ 
 & $0.234 + [0.008+0.0i+ \ldots]$ 
 & $0.99 \zeta^{B\pi}+ \{ 0.79, 0.74\} \zeta_J^{B\pi}$
 &  $T_c=0.16 \pm 0.02$ \\
 &\hspace{0.2cm} $-(0.016 + 0.001 X_H + \ldots)$ 
 &\hspace{0.2cm}  $-(0.018 + 0.003 X_H+ \ldots)$
  & \\
\end{tabular}
\end{ruledtabular}
\caption{\label{table}
Numerical results for the $B\to \pi\pi$ amplitudes. The first two columns are
QCDF results organized so terms with $\alpha_s(\sqrt{E\Lambda})$ are included at
lowest order, and using two scenarios for the nonperturbative input parameters
from~\cite{BN}. The numbers in the square brackets are the ${\cal
  O}(\alpha_s(m_b))$ corrections, and the numbers in round brackets are so-called
``chirally enhanced'' $\Lambda/m_b$ corrections. Unknown
$\alpha_s(m_b)$ and $\Lambda/m_b$ corrections that contribute at the same order
are indicated by ellipses. The third column is the LO SCET result in terms of
nonperturbative parameters. The last column is the experimental data from an
isospin analysis using $\gamma=64^\circ$ and $|V_{ub}|=3.9\times 10^{-3}$. The
results for all amplitudes have a common prefactor $N_\pi = G_F m_B^2
f_\pi/\sqrt2$ removed.
}
\end{table*}

Assuming isospin symmetry and neglecting the electroweak penguins, the 
$B\to\pi\pi$ amplitudes can be written as
\begin{eqnarray}\label{b2pipi}
A(\bar B^0\to \pi^+\pi^-) &=& \lambda_u^{(d)}\, T_c + 
\lambda_c^{(d)}\, P \,, \nn\\
\nonumber
A(\bar B^0\to \pi^0\pi^0) &=& \lambda_u^{(d)}\, T_n 
- \lambda_c^{(d)}\, P \,,\\
\sqrt2 A(B^-\to \pi^0\pi^-) &=& \lambda_u^{(d)}\, T  \,.
\end{eqnarray}
This contains five independent hadronic parameters which can be extracted
from an isospin analysis. Using the world averages of the data \cite{pipidata} 
and setting $\gamma = 64^\circ$ one finds the results in the last column of
Table~\ref{table}, and
\begin{eqnarray}
  |T_n|/N_\pi &=&\Big\{
   \parbox{2.2cm}{$0.15\pm 0.02 ({\rm I})$, $0.18\pm 0.02 ({\rm II}) $} \,,\\
  P/N_\pi &=& (-0.024\pm 0.007) +(0.021 \pm 0.007 )i \,.  \nn
\end{eqnarray}
Note that our power counting for $A_{c\bar c}^{\pi\pi}$ in Eq.~(\ref{Acc}) gives
complex values of a similar size, $|P|/N_\pi \sim \alpha_s(2m_c) v T_c/N_\pi\sim
0.03$, where we took $f(2m_c/m_b)\sim 1$.

Predictions in QCDF involve an expansion at the intermediate scale and use
sum-rule calculations for $\phi_B(k^+)$ since this hadronic function is not
known from data.  Our approach was to instead fit the parameter $\zeta_J^{B\pi}$
to the nonleptonic data and using $\langle x^{-1}\rangle_\pi=3.0$ which falls
within the range, $3.2\pm 0.4$, preferred by fits to the $\gamma^*\gamma\to\pi^0$
data\cite{gampi0}. We also assumed that $\alpha_s(m_b)$ corrections will be of a
similar size to neglected power corrections. Numerical justification for this is
discussed below.

Factorization formulas like Eq.~(\ref{Apipi}) express the amplitudes $T$ and
$T_c$ at leading order in $1/m_b$ in terms of the nonperturbative parameters
$\zeta^{B\pi}$ and $\zeta_J^{B\pi}(x)$.  Using the leading order SCET relations
from~\cite{bprs} we find
\begin{eqnarray}\label{SCETLO}
\left. \zeta^{B \pi} \right|_{\gamma = 64^\circ}^{\rm leading} &=& 
(0.08 \pm 0.03) \left( \frac{3.9 \times 10^{-3}}{|V_{ub}|} \right) \,, \nn\\
\left. \zeta_J^{B \pi} \right|_{\gamma = 64^\circ}^{\rm leading} &=& 
(0.10 \pm 0.02) \left( \frac{3.9 \times 10^{-3}}{|V_{ub}|} \right) \,.
\end{eqnarray}
These can be used to predict the $B\to \pi$ form factor at $q^2=0$ as $ f_+(0) =
\zeta^{B\pi} + \zeta_J^{B\pi}$.  In Fig.~\ref{fig:ff} we show how results for
our extraction of the $\zeta$'s and $f_+(0)$ depend on the input value of
$\gamma$ (with normalization taken using the value of $|V_{ub}|$ preferred by
current inclusive fits~\cite{newVub}). Note that the smaller value of $f_+(0)$
favored from our analysis would increase the value of $|V_{ub}|$ from exclusive
decays, perhaps bringing it in line with the inclusive analyses.

Working at tree level in the jet function, $ f_+(0) = \zeta^{B\pi} +
\zeta_J^{B\pi}$ and the $\zeta_J$ parameter is given by
\begin{eqnarray}\label{QCDF}
  \zeta_J^{B\pi} = \frac{\pi \alpha_s(\mu_{\rm int}) C_F}{N_c} 
    \frac{f_B f_\pi}{m_B \lambda_B} \langle y^{-1} \rangle_\pi\,.
\end{eqnarray}
With the input parameters adopted in QCDF~\cite{BBNS,BN} at leading order in
$\alpha_s$ and $\Lambda/m_b$ one can find values for the SCET parameters.  In
the default scenario of~\cite{BN} we find: $\zeta^{B\pi}=0.26$,
$\zeta_J^{B\pi}=0.02$, while in their scenario 2: $\zeta^{B\pi}=0.20$,
$\zeta_J^{B\pi}=0.05$, which are quite different from Eq.~(\ref{SCETLO}).  There
are two possible explanations for this disagreement: i) higher order
perturbative and power corrections are important; ii) some of the hadronic input
parameters used in QCDF are not supported by the data. The authors
of~\cite{BBNSComment} take the first point of view and argue that there are
large known perturbative and power corrections to the leading order result.

To compare with QCDF we have calculated the hadronic parameters $T,T_c$ using
the analysis in~\cite{BBNS,BN}. The results are shown in Table~\ref{table} for
two sets of input parameters from~\cite{BN}, their default scenario and their S2
scenario. We have organized the terms according to the expansion advocated in
Ref.~\cite{bprs} with $\alpha_s(\sqrt{E\Lambda})$ terms included at LO. The
first line in the table shows the LO terms with $\alpha_s(m_b)$ corrections in
square brackets, and the second line shows $\Lambda/m_b$ corrections. We have
dropped other $\Lambda/m_b$ terms which contribute $\lesssim 6\times 10^{-3}
N_\pi$ in the amplitudes $T$, $T_c$ for default inputs. In contrast to
Ref.~\cite{BN}, we evaluate all full theory $C_i$'s at $\mu=m_b$, since below
this scale the running does not follow from the usual anomalous dimensions of
the electroweak Hamiltonian.  Note that $\int dx \phi_\pi(x)/x =3.3$ in the
default scenario and $\int dx \phi_\pi(x)/x =4.2$ in the S2 scenario.  We give
the corresponding LO SCET results for these two cases in the table.

From Table~\ref{table} the $\alpha_s(m_b)$ perturbative corrections amount to a
$\lesssim 10\%$ shift in the leading order results for $T, T_c$. The ``known''
non-perturbative corrections with no $X_H$ factor are $\lesssim 10\%$.
Non-perturbative corrections proportional to divergent convolutions in QCDF are
$\lesssim 10\%$ for the canonical choice $X_H=2.4$ from Ref.~\cite{BN}.  All of
these are of a similar size to the $10-20\%$ corrections we expect from other
unknown power corrections. Thus, with the expansion in~\cite{bprs} the model
parameters used in QCDF support the claim that the leading order extraction of
$\zeta^{B\pi}$ and $\zeta_J^{B\pi}$ from $T,T_c$ is good to $\sim 20\%$. The
reason Ref.~\cite{BBNSComment} found that power corrections change
$\zeta_J^{B\pi}$ substantially is that their input parameters give a LO result
that is small, numerically of a similar size to a typical power correction.

In conclusion, all the perturbative and power corrections which are truly known
give rise to small shifts in our LO analysis. The criticisms of
Ref.~\cite{BBNSComment} about our work are based on guesses about the size of
hadronic parameters and the size of power corrections and thus in our opinion
not reliable. Further work is required to get a better understanding of power
corrections for non-leptonic $B$ decays.  Some interesting work in this
direction has been done recently in Ref.~\cite{FH}.

\section{Conclusions}

In this letter we discussed the differences between the QCDF approach and SCET
approach to factorization in nonleptonic $B\to \pi\pi$ decays, expanding on the
points already made in \cite{bprs} and addressing the criticism
in~\cite{BBNSComment}.  We also commented on recent SCET work related to these
points, which appeared after the publication of \cite{bprs}.

We addressed the main points made in Ref.~\cite{BBNSComment}. Formally, SCET
tells us that either there are missing $\alpha_s(m_b)$ corrections
in~\cite{BBNS,BN} (for $\zeta^{B\pi}\sim \zeta_J^{B\pi}$), or the QCDF counting
which treats $\alpha_s(m_b)\sim \alpha_s(\sqrt{E\Lambda})$ relies on
$\zeta^{B\pi}\gg \zeta_J^{B\pi}$. Avoiding perturbation theory at the
intermediate scale $\mu_{int} \simeq \sqrt{\lqcd m_b}$ might seem to introduce
more nonperturbative $\zeta_J(z)$ functions than expanding in
$\alpha_s(\mu_{int})$.  However, when restricted to the subset of nonleptonic
and semileptonic B decays into pions, this amounts simply to trading one unknown
function for another ($\phi_B(k_+)$ vs.  $\zeta_J^{B\pi}(z)$).  Contrary to the
claims made in~\cite{BBNSComment}, we still find a complex $A_{c\bar
  c}^{\pi\pi}/A_{LO}^{\pi\pi}\sim \alpha_s(2m_c) v$, indicating that long
distance charm penguin contractions can be numerically significant.  Finally, we
show that our phenomenological analysis of $B\to \pi\pi$ data, and the
determination of the two hadronic parameters $\zeta^{B\pi}$ and $\zeta_J^{B\pi}$
remain correct when known perturbative and non-perturbative corrections are
estimated as in Ref.~\cite{BN}. The largest corrections actually come from
unknown power suppressed terms, but are still within our error estimate.  We
leave it to the reader to assess the relative importance of the agreements and
disagreements.

This work was supported in part by the DOE under DE-FG03-92ER40701,
DOE-ER-40682-143, DEAC02-6CH03000, and the cooperative research agreement
DF-FC02-94ER40818, and by a DOE Outstanding Junior Investigator award and Alfred
P.~Sloan Fellowship (I.S.).

\vspace{-.4cm}

\begin{appendix}
\section{
Scaling of $A_{c\bar c}^{\pi\pi}$}
\label{appendix}

In this appendix we derive the scaling in Eq.~(\ref{Acc}). First consider limit
i). Above the b-mass we have the operator $(\bar c \Gamma b)(\bar d \Gamma c)$.
Next we integrate out the scale $m_b$ and match this operator onto one with
massive collinear charm quarks (to ensure they are moving close together, i.e.
have invariant mass $\sim 4 m_c^2$) and a collinear light $d$ quark,
\begin{eqnarray} \label{Oprod0}
  O^{\rm prod} = \big[ \bar \xi^{(d)}_{\bn,\omega_1} \Gamma' T^A h_v^{(b)} \big]
   \big[ \bar \xi^{(c)}_{n'} \Gamma  T^A \xi^{(c)}_{n'} \big]\,,
\end{eqnarray}
Here $\omega_1=m_b u$ and the massive collinear charm quarks have $\bn'\cdot
p\sim m_b$ and $p_\perp \sim m_c$ with the Lagrangian from~\cite{LLW} (we omit
all Wilson lines and other factors that are irrelevant to the power counting).
These \SCETa fields have momenta $p^2\sim m_b\Lambda$ or $m_c^2$ which we treat
as the same size.  The collinear expansion parameters are
$\lambda=\sqrt{\Lambda/m_b}$ and $\lambda_c=m_c/m_b$, and $\Gamma = \bnslash$
will not contribute, but mass insertions~\cite{Ira} will contribute, such as
$\Gamma= \bnslash \gamma_\perp m_c/(i\bn\mcdot D)\sim \lambda_c$ as seen below.
The operator therefore scales as $O^{\rm prod} \sim m_b^6 \lambda^4
\lambda_c^3$.

Next we integrate out the scale $m_c$. The scale $\sqrt{E_\pi\Lambda}$ is close
to $m_c$ and can be integrated out at the same time, but the factors generated
from doing this are the same as those from the $O^{(0)}_{\rm SCET_I}$ which
occur in the non-charm contributions and so do not effect the relative scaling.
Removing $m_c$ in $O^{\rm prod}$ requires matching the charm fields onto NRQCD
fields $\eta$, $\chi$.  There is an operator $O^{\rm ann}_{\rm I}$ which
annihilates the charm in the boosted frame, and at tree level comes with a
$1/(4m_c^2)$ prefactor from integrating out a single gluon\footnote{We use BR to
  denote the fact that the matrix elements of these fields need to be boosted to
  the $B$ rest frame. The $v$ scaling is assigned from the CM frame, but aside
  from $L$ is not affected by the boost since there is an invariant that can be
  formed for the resulting matrix element without kinematic factors.}
\begin{eqnarray}  \label{CCpair}
O^{\rm prod}_{\rm I}(0) &=& 
  \left[\bar{\xi}_{\bn,\omega_1}\Gamma T^A h_v\right]
  \big[\eta^\dagger  T^A (\sigma\cdot L) \chi \big]_{\rm BR} \,, \\
O^{\rm ann}_{\rm I}(x) &=& \Big[ \frac{1}{4m_c^2}
   \chi^\dagger  T^A \sigma_\perp \eta \Big]_{\rm BR}
  \left[ \bar\xi_{n,\omega_3} \gamma_\perp T^A \xi_{\bn,\omega_2} \right] \,.\nn
\end{eqnarray}
The boost matrix $L$~\cite{Braaten} depends on whether the $\sigma$ is $\perp$
or longitudinal. The annihilation operator is $\perp$ if $\alpha_s(2m_c)$ is a
good expansion parameter, so we take $\sigma_\perp$ in $O^{\rm prod}_{\rm I}$ in
which case $L\sim 1$. Since $[\eta^\dagger T^A \chi] \sim m_c^3 v^3$ this
reproduces the $m_c^3$ for Eq.~(\ref{Oprod0}).  Power counting gives
\begin{eqnarray}
  O_{\rm I}^{\rm prod} &\sim& \big(m_b^3 \lambda^4\big) 
  \Big( m_c^3 v^3  \Big)
  \,, \  \nn\\
  O^{\rm ann}_{\rm I} &\sim& \Big( \frac{m_c^3 v^3}{4m_c^2} \Big) 
  \big( m_b^3 \lambda^2 \big) \,.
\end{eqnarray}
Label momentum conservation in $O_{\rm I}^{\rm prod}$ implies that the $c\bar c$ total
momentum has $q^2\equiv m_b^2 \bar u = 4 m_c^2$ where $\bar u=1-u$. This gives a
delta function in the Wilson coefficient, 
\begin{eqnarray} \label{C1prod}
 C_{\rm I}^{\rm prod} \simeq  
    \delta\Big(\bar u - \frac{4m_c^2}{m_b^2} \Big) \,.
\end{eqnarray}
The $\delta$-function is expected~\cite{FH}, and is analogous to Eq.~(3.13)
of~\cite{BR} for the factorization formula used in production of energetic
$c\bar c$ state's, whose hadronization is governed by NRQCD. Any
$\delta$-function for $O_{\rm I}^{\rm ann}$ just ensures overall momentum
conservation and can be omitted, so we count $C_{\rm I}^{\rm ann}\simeq
\alpha_s(2m_c)$.

If we now consider the time ordered product capturing the NRQCD region we have
\begin{eqnarray} \label{scaling}
  && C_{\rm I}^{\rm prod} C_{\rm I}^{\rm ann} 
  \int\!\! d^4x\:  T\big[ O_{\rm I}^{\rm prod}(0) O^{\rm ann}_{\rm I}(x) \big]
  \\
  && \sim C_{\rm I}^{\rm prod} C_{\rm I}^{\rm ann} 
   (m_c^{-4} v^{-5}) (m_b^3 m_c^3 \lambda^4  v^3) (m_b^3 m_c \lambda^2 v^3)
  \nn\\
  &&\sim (m_b^6 \lambda^6)\, C_1^{\rm prod} C_1^{\rm ann}  
   \: v \nn\\
 &&  \sim \Big[ O^{(0)}_{\rm SCET_I} \Big] \Big\{  \alpha_s(2m_c) 
   \delta\Big(\bar u - \frac{4m_c^2}{m_b^2}\Big) \: v \Big\}
   \,.\nn
\end{eqnarray}
Our result for the relative scaling of the $B\to\pi\pi$ amplitudes in limit i),
$A_{c\bar c}^{\pi\pi}/A^{\pi\pi}_{LO}$, is in curly brackets, and gives
Eq.~(\ref{Acc}) (once it is integrated over the $\bar u$ dependence from the
matrix element).  Note that the scaling of NRQCD fields together with
(\ref{C1prod}) account for the ``phase space suppression'' factors, and that
there is an enhancement from the charm propagators being close to their mass
shell.  We have not been careful to factorize the usoft gluons from the
collinear fields etc., so (\ref{scaling}) only indicates the scaling and can not
be used to determine the factorization or final operator in \SCETb.  Integrating
over a collinear matrix element that depends on $\bar u$, $\psi(\bar u)$, will
induce additional dependence on $2m_c/m_b$, so in limit (i) we have to make some
assumption about the form of the wavefunction (which we have not proven) to
quote a scaling.  Taking the form $\psi(\bar u) \sim 6 \bar u (1-\bar u)$ with
$4m_c^2/m_b^2\simeq 0.44$ unexpanded gives $A_{c\bar
  c}^{\pi\pi}/A_{\pi^+\pi^-}\sim 1.5\, v\, \alpha_s(2m_c)$.

In limit (ii) we can integrate out $m_b$ and $m_c$ simultaneously, and the
calculation above of $f(2m_c/m_b)$ will change. Inserting $f(2m_c/m_b)\sim 1$ in
Eq.~(\ref{Acc}) gives $A_{c\bar c}^{\pi\pi}/A^{\pi\pi}_{LO}\sim v\, \pi
\alpha_s(2m_c)$, which is the result from~\cite{bprs}. In this case there are
clearly (u)soft gluons that couple, the $c\bar c$, the soft $b$, and soft
spectator.

Note that we have only considered a schematic argument for the scaling of these
NRQCD $c\bar c$ contributions rather than deriving a factorization theorem, so
they are not at the same level of rigor as (\ref{Apipi}). It is possible that a
full consideration of the factorization for $A_{c\bar c}^{\pi\pi}$ might uncover
an additional subtelty which changes our conclusions. Obviously one can not use
the scaling arguments to fix values for $A_{c\bar c}^{\pi\pi}$ so we instead
used experimental data for the penguin, $P$. Note that this method of analysis 
is also appropriate if the power expansion for $P$ does not converge.

\end{appendix}

\end{document}